\documentstyle[12pt,epsfig]{article}

\begin{document}

\begin{titlepage}
\hfill       OHSTPY-HEP-T-99-019\\
\mbox{ } \hfill      December 1999 \\

\vspace{1.0cm}

\begin{center}

{\Large\bf An analysis of a Heavy Gluino LSP at CDF :}

\vspace{.5cm} 

{\Large\bf The Heavy Gluino Window}

\vspace{2cm}

{\large\bf Arash Maf\hspace{1pt}i and Stuart Raby}

\bigskip

{\em Department of Physics\\ The Ohio State University\\ 174 W. 18th Ave.\\
Columbus, Ohio  43210}

\vspace{1cm}

\end{center}

\abstract
{
In this paper we consider a heavy gluino to be the lightest supersymmetric particle [LSP]. We investigate the limits on the mass of a heavy gluino LSP, using the searches for excess events in the jets plus missing momentum channel in Run I. The neutral and charged R-hadrons, containing a heavy gluino LSP, have distinct signatures at the Fermilab Tevatron. The range of excluded gluino masses depends on whether the R-hadron is charged or neutral and the amount of energy deposited in the hadronic calorimeter. The latter depends on the energy loss per collision in the calorimeter and the number of collisions; where both quantities require a model for R-hadron- Nucleon scattering.   We show how the excluded range of gluino mass depends on these parameters.  We find that gluinos with mass in the range between $\sim 35$ GeV and $\sim 115$ GeV are excluded by CDF Run I data.  Combined with previous results of Baer et al., which use LEP data to exclude the range 3 - 22$\sim$25 GeV, our result demonstrates that an allowed window for a heavy gluino with mass between 25 and 35 GeV is quite robust.  Finally we discuss the relevant differences of our analysis of Tevatron data to that of Baer et al.      
}

\end{titlepage}

\section{INTRODUCTION}
\label{sec:intro}

Supersymmetric theories are the most promising candidates for new physics beyond the standard model.  Since supersymmetry [SUSY] is not observed at low energies, it must be broken. Two interesting classes of models for SUSY breaking are minimal supergravity \cite{sugra}(also known as the constrained minimal SUSY standard model [CMSSM] \cite{cmssm}) and {\em minimal} gauge-mediated SUSY breaking [GMSB] \cite{gmsb}.  In both cases gauginos are assumed to have a common mass at the grand unification [GUT] scale \cite{drw}. Since gaugino masses, at one loop, are proportional to their corresponding coupling constants, we naturally expect {\em gluinos} to be the heaviest gauginos at low energies.  However, ref. \cite{Raby1,Raby2} has shown it is possible (and even natural) to build phenomenologically acceptable GMSB models with a heavy gluino as the LSP. \footnote{This is not the only model which may lead to a gluino LSP; see for example, the so-called O-II string model discussed in ref.~\cite{string} or another GMSB model discussed in ref.~\cite{nandi}.}
  In these GUT models, gaugino masses do NOT unify at the GUT scale. 

In GMSB models, supersymmetry is broken in a hidden sector and transmitted to the visible sector via messengers carrying standard model gauge interactions.  The masses of gauginos arise via one loop diagrams containing messenger fields with appropriate quantum numbers.  The minimal messenger sector includes color triplets, giving mass to gluinos and squarks, and weak doublets, giving mass to charginos and neutralinos as well as squarks and sleptons.   These minimal messengers form complete five dimensional representations of SU(5); thus preserving GUT predictions.  If the color triplet and weak doublet messengers are degenerate, then gaugino masses $\tilde m_i, i= 1,2,3$ satisfy $\tilde m_i/\alpha_i = constant$ at the messenger scale.  If, on the other hand, the color triplet messengers are heavier than their weak doublet partners, gluino masses are suppressed. In the model of ref. \cite{Raby1,Raby2}, it was argued that the latter possibility is ``natural" in an SO(10) SUSY GUT with minimal messenger sector.  In this case, the Higgs and messenger fields both belong to the $5 + \bar{5}$ of $SU(5)$ (or a 10 of $SO(10)$).    Since they have identical quantum numbers it is natural for them to mix.  Moreover the natural mass scale for real SU(5) representations is of order the GUT scale.  Recall, in any SUSY GUT, Higgs triplet and doublet masses must be split in order to avoid rapid proton decay.  It was shown in a simple example that the combination of Higgs-messenger mixing and doublet-triplet splitting leads to color triplet Higgs and messengers with mass at the GUT scale ($M_G$) and lighter doublet messengers with mass at a scale $M < M_G$.  The Higgs doublets remain massless.  $M$ is identified as the messenger scale.  In this scenario the gluino is the LSP.  Assuming a conserved R-parity, such a gluino is stable.

A natural framework for the above scenario is obtained in $SO(10)$ with the Dimopoulos-Wilczek mechanism~\cite{Dimopoulos} for doublet-triplet splitting.   This theory requires two $10$s, the Higgs $10$ dimensional multiplet and an auxiliary $10$ field.  The Higgs and auxiliary fields mix; and if the auxiliary $10$ feels SUSY breaking at tree level we have GMSB with the auxiliary $10$ identified as the messengers \cite{Raby1}.  Since $M$ is both the messenger scale and the R symmetry breaking scale, the gluino mass is suppressed by $(M/M_G)^2$ compared to other gaugino masses. Note, in this theory, as shown in ref. ~\cite{babubarr}, $M/M_G < 0.1$ is also sufficient to suppress the baryon number violating nucleon decay rates. 

Given the possibility of having a gluino LSP it is important to consider the existing limits.\footnote{There are also cosmological constraints on a gluino LSP\cite{cosmological}.  These constraints, if correct, would require the gluino to be the NLSP and to decay, for example, into a gravitino (such as discussed in refs.~\cite{Raby2,Baer}) or into a neutralino as in \cite{Farrar}.}  In a much studied class of models~\cite{Farrar} gauginos are massless at tree level obtaining their mass radiatively at low energies.  In these {\em radiative} gaugino models the lightest neutralino is the LSP with a gluino NLSP (next to the lightest superparticle) with mass of order a couple of GeV or less. These models provide candidates for dark matter ~\cite{chung} and UHECRons ~\cite{albuquerque}, the particles responsible for the Ultra High Energy Cosmic Rays observed by cosmic ray shower detectors.   There are two classes of experimental constraints on these models.  The first class looks for the decay of the gluino into a quark- anti-quark pair and the LSP.  Such a search by KTeV ~\cite{ktev} has ruled out these models in the entire region of parameter space consistent with the dark matter and UHECRon solutions.   The second class of constraints is valid even if the gluino were the LSP and thus stable.   For example,  beam dump experiments at Fermilab, as well as, searches for the decay $\Upsilon \rightarrow \eta_{\tilde g} + \gamma$ rule out gluinos with mass in the range $\sim 2 - 4$ GeV.  In addition, gluinos in the entire light gluino window $1 - 5$ GeV are ruled out by searches for dijet events at Fermilab for squark masses roughly in the range $150 - 600$ GeV~\cite{hewett}.  Finally the analysis of the running of $\alpha_s$ from $m_\tau$ to $M_Z$ combined with multijet angular correlations on the Z has lead the authors of ~\cite{Csikor} to claim that the entire light gluino window is ruled out.   The analysis of the multijet angular variables however has large uncertainties as noted in ~\cite{Farrar1}.   If we neglect this part of the analysis, only using the running of $\alpha_s$, then the light gluino window is ruled out only at 80\% CL ~\cite{Csikor}.   In summary, the light gluino window may still be viable for stable gluinos.  

In this paper, however, we only consider models with a heavy gluino LSP with mass greater than 5 GeV.    In fact, as has been discussed in ~\cite{Raby1},  all previous searches ruling out heavy gluinos in the MSSM (with mass greater than 5 GeV) use the jets plus missing momentum signature coming from the sequential decay of a gluino into $q + \bar q$ + the LSP (typically the lightest neutralino).    However, if the gluino is the LSP and thus stable (assuming a conserved R-parity) then all previous limits on a heavy gluino, and most searches for SUSY, must be re-evaluated.  Hence it is important to study a heavy gluino LSP (in the {\em heavy gluino window}). 

Current experimental data can greatly constrain these models ~\cite{Raby1,Raby2}, \cite{Starkman} - \cite{Gunion}. 
An important new study of the heavy gluino LSP has been carried out by Baer, Cheung and Gunion [BCG] ~\cite{Baer}.  BCG have significantly constrained the heavy gluino window, using both OPAL and CDF data; their results are presented below. An important parameter in their analysis is $P$ or $1 - P$, the probability for a gluino to fragment into a charged or a neutral R-hadron.\footnote{An R-hadron denotes the color singlet bound state of a gluino with gluon or light quark constituents.}  The same parameter, in their analysis, is used for the probability of the final state R-hadron in an R-hadron - Nucleon collision to be charged or neutral.

\begin{itemize}
\item  {\em Using OPAL data~\cite{opal} for $e^+ e^- \rightarrow Z \rightarrow N_2 N_1$ with $N_2 \rightarrow N_1 + q + \bar q$ where the $N_2 (N_1)$ neutralino is the NLSP (LSP) in the MSSM:}  This process was analyzed by OPAL for 2, 3 or 4 jets + missing momentum. Analyzed in terms of the process $ e^+ e^- \rightarrow q \bar q \tilde g \tilde g$ BCG find that gluino masses in the range 3 - 22-25 GeV are ruled out.   This result is independent of the probability $P$.
\item {\em Using CDF data for jets + missing momentum ~\cite{Exper}:} BCG find that gluinos with mass in the range 25 to 130 - 150 GeV are ruled out for all values of $P$ in the range $0 \le P < 3/4 $.  Combined with the OPAL data, this excludes a heavy gluino LSP in the range from 3 to 130 - 150 GeV.  However for values of $P \geq 3/4  $ and large hadronic scattering length, they find an allowed heavy gluino window from 23 - 50 GeV.  \end{itemize}

It is the goal of this paper to better understand the sensitivity in the BCG analysis to the probability parameter $P$ and hence the existence of the heavy gluino window.  Thus we re-analyse the CDF data using an entirely different model for R-hadron - Nucleon scattering.  As in the analysis of BCG, we first assume the gluino LSP is heavy and that all other supersymmetric particles are substantially heavier than the gluino.  We then check the sensitivity of our results to lighter squarks.  Our results are found in sec. \ref{sec:result}.   {\em The basic conclusion is that there exists a heavy gluino window in the mass range from 25 - 35 GeV.}  

The paper is organized as follows.  We review the processes responsible for R-hadron production and fragmentation at the Tevatron in sec. \ref{sec:prod} and for detection at CDF in sec. \ref{sec:const}.  In particular, the Regge model used for R-hadron - Nucleon scattering is found in sec. \ref{sec:regge}.  In sec. \ref{sec:how} we present the details of our analysis, i.e. evaluating missing momentum;  ionization energy loss, and muon identification.   Our results are presented in sec. \ref{sec:result} and the conclusions are given in sec. \ref{sec:conclude}.

\section{A HEAVY GLUINO LSP AT THE TEVATRON}
\subsection{Production and Fragmentation}
\label{sec:prod}

\begin{figure}
\scalebox{1.0}[1.0]
{
\includegraphics{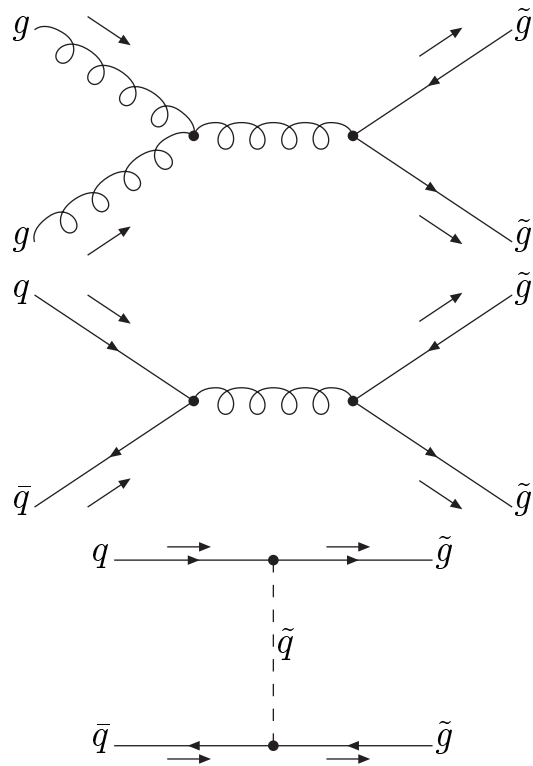}
}
\caption{Diagrams representing gluino pair production
in $p-\bar{p}$ collisions.}
\label{f:collision}
\end{figure}

If gluinos are light enough, they can be produced in proton-antiproton collisions at the Tevatron. The relevant Feynman diagrams are shown in Fig.\ \ref{f:collision}.

Gluinos are color octets; hence gluinos cannot exist in isolation.  They form 
colorless bound states such as $R_0 = \tilde{g}g$ (gluino-gluon bound state, an isospin singlet) or $\tilde \rho = \tilde{g} (q\bar{q})_{(I=1)}$ (gluino-quark-antiquark bound state with isospin 1) with ($\tilde{\rho}^+ = (\tilde{g}u\bar{d})$, $\tilde{\rho}^0 = (\tilde{g}u\bar{u}-\tilde{g}d\bar{d})/\sqrt{2}$, $\tilde{\rho}^- = (\tilde{g}d\bar{u})$).   In our analysis, we assume that these four states are the lightest bound states of the gluino.   Together, they are called R-hadrons.  As we see, two of these four R-hadrons are electrically charged and two are neutral.  We also assume that any other bound state of the gluino, when produced, rapidly decays into one of these four R-hadrons.

If $m_{\tilde \rho} > m_{R_0} + m_\pi$ then the decay $\tilde \rho \rightarrow R_0 \pi$ will occur on a strong interaction time scale of order $10^{-23}$ s.  Otherwise $\tilde \rho^\pm \rightarrow R_0 l^\pm \nu$ or
$\tilde \rho^0 \rightarrow R_0 \gamma$ occur via weak or electromagnetic interactions.  

For  $\tilde{\rho}^+$ we obtain the decay rate 
\begin{eqnarray}
\Gamma(\tilde{\rho}^+ \to R^0 e^+ \nu)= &
\frac{\Gamma(\tilde{\rho}^+ \to R^0 e^+ \nu)}{\Gamma(\rho\to e^+e^-)}
\times \Gamma(\rho\to e^+e^-) &
\end{eqnarray}  
where the ratio
\begin{eqnarray}
\hat{r}^+  \equiv & \frac{\Gamma(\tilde{\rho}^+ \to R^0 e^+ \nu)}{\Gamma(\rho\to e^+e^-)}= & \frac{2^{9}}{15}\frac{G^2_F \alpha_s m^2_u \Delta M^4}{\alpha^2 M^2_R} ,
\end{eqnarray}  
$\Delta M = m_{\tilde \rho^\pm} - m_{R_0}$, and  $M_R$ and  $m_u$ is the R-hadron and up quark mass, respectively.  
Using the experimental value 
\begin{eqnarray}
\Gamma(\rho\to e^+e^-) = 0.00675 \;\; {\rm MeV}
\end{eqnarray}  
we find the lifetime $\tau_{\tilde{\rho}^+} = 10^{-7} s$  for $ M_R=50$ GeV  and $\Delta M=140$ MeV or $c \tau_{\tilde \rho^\pm} \sim 30$ m.  

Similarly for $\tilde \rho^0$ we have
\begin{eqnarray}
\Gamma(\tilde{\rho}^0\to R^0+\gamma)= &
\frac{\Gamma(\tilde{\rho}^0\to R^0+\gamma)}{\Gamma(\rho\to e^+e^-)}
\times \Gamma(\rho\to e^+e^-) &
  \end{eqnarray}  
and we obtain~\footnote{For this calculation we use the analysis leading to equation (4) ~\cite{chanowitz} on the radiative decay $J/\Psi \rightarrow \gamma gg$.}
\begin{eqnarray}
\hat{r}^0  \equiv \frac{\Gamma(\tilde{\rho}^0\to \gamma+g+g)}{\Gamma(\rho\to e^+e^-)}=
\frac{8}{45}\times \frac{(\pi^2-9)}{\pi}
\frac{\alpha_s^2}{\alpha}\frac{\Delta M^4}{M^2_R m^2_u}
\end{eqnarray} 
Using the same numerical values as above, we find the lifetime 
$\tau_{\tilde{\rho}^0} \sim 10^{-14} s$ or $c \tau_{\tilde{\rho}^0} \sim 0.0003$ cm.  Note for smaller values of $\Delta M$ the lifetime quickly grows.    
For $\Delta M = 5$ MeV, we have $c \tau_{\tilde{\rho}^0} \sim 200$ cm.

Gluinos produced in a proton-antiproton collision fragment into one of these bound states.  The probabilities of fragmenting into $R_0$, $\tilde{\rho}^+$, $\tilde{\rho}^0$ and $\tilde{\rho}^-$ are denoted by
$P_{R^0}$, $P_{\rho^+}$, $P_{\rho^0}$ and $P_{\rho^-}$, respectively with  $P_{R^0}+P_{\rho^+}+P_{\rho^0}+P_{\rho^-}=1$.  If a particle lifetime is too short for the particle to multiple scatter in the detector, then the effective fragmentation probability for that particle is zero (see discussion of $\lambda_T$ in sec. \ref{sec:lambdat}).  For example, if $m_{\tilde \rho} > m_{R_0} + m_\pi$ then we have $P_{R_0} = 1$.  On the other hand for  $m_{\tilde \rho} < m_{R_0} + m_\pi$ and taking $\Delta M=140$ MeV and $M=50$ GeV in the equation above we find  $c \tau_{\tilde \rho^0} \sim 3 \times 10^{-4}$ cm giving   $P_{\rho^0} \approx 0$.  If all non-vanishing probabilities are then taken approximately equal we find the probability $P$ (as in BCG) for producing a charged R-hadron given by $P \sim 2/3$.\footnote{Note, in our analysis these probabilities are only used for the initial fragmentation.} If, however, we take $\Delta M = 5$ MeV instead, then all four probabilities are effectively non-zero and $P \sim 1/2 $.  In general, the effective fragmentation probabilities depend on $\Delta M$, $M$ and the relativistic $\gamma$ factor.  In our analysis, however, we take these fragmentation probabilities to be free parameters and we test the sensitivity of our results to changes in them.

\subsection{CONSTRAINTS FROM CDF}
\label{sec:const}

To constrain the mass of the heavy gluino LSP we use, following the analysis of Baer {\it et al.} \cite{Baer}, CDF data for jets+${\not p}_T$ \cite{Exper}.  This data has been used by CDF to put limits on gluino and squark masses.   In the MSSM a heavy squark or gluino typically decays sequentially producing several jets plus missing momentum carried away by the LSP.  In addition, in order to reduce the standard model background for the jets +${\not p}_T$ signature, CDF cuts all jets containing a charged lepton.   

In our case the lightest R-hadron is the LSP and since it is a hadron (in contrast to the CMSSM) it will deposit some of its energy in the hadronic calorimeter.   In addition,  the R-hadron's charge may fluctuate due to hadronic collisions.   Thus as emphasized by BCG it is important to analyze in detail how an R-hadron is observed in the detector.  {\em In fact it was shown that some of the time a charged R-hadron may be identified as a muon.}

   The important components of the CDF detector (for our analysis) are the central tracker, hadronic calorimeter, the inner and outer muon chambers and the un-instrumented piece of iron, located between the muon chambers. Consider how an R-hadron is observed.  A gluino fragments into one of the four stable R-hadrons.\footnote{Note, typically the R-hadron carries most of the gluino momentum ( $\sim$ 99.4\% for a 50 GeV gluino); leaving little or none for associated pions or other hadrons.  Thus a gluino jet often contains only the single R-hadron and nothing else.}  If the R-hadron is charged, it leaves a track in the central tracker.  Then the R-hadron enters the hadronic calorimeter (layers of iron) where it scatters off the nucleons a few times before it is stopped or, if not, it enters the inner muon chamber.\footnote{As discussed in ref. \cite{Drees}, it is reasonable to take the individual nucleon as the target for values of $1 > |t| \geq 0.01$ GeV$^2$.  For smaller values of $|t|$ the entire nucleus would be treated coherently and for larger $|t|$ individual quarks would be the scatterers.}  After each hadronic collision, the R-hadron can change into any other possible ``stable" gluino bound state, i.e. any of the four R-hadrons with the probability determined by the appropriate scattering cross section.  A charged R-hadron also deposits ionization energy in the hadronic calorimeter.  Note, if the R-hadron exiting the hadronic calorimeter is charged, it is detected in the muon chambers.  In addition, between the first and second chambers it can lose energy and also the charge of the R-hadron can change due to an hadronic interaction inside the un-instrumented iron.  The energy loss will again be both in the form of hadronic energy loss and ionization (if it is charged).  However, this energy loss is not detected.  

In order to determine the amount of energy deposited in the hadronic calorimeter and to model the charge fluctuations in the detector, we need a model for R-hadron scattering.  The magnitude of the total cross section $\sigma_T$ determines the scattering length $\lambda_T  \propto 1/\sigma_T$.  The angular dependence of the differential scattering cross section determines the mean energy loss per collision $\Delta E$.     Finally the differential cross sections determine the probability for charge fluctuations.   In these details reside the main differences between our analysis and that of BCG.

\subsubsection{The total cross section and the interaction length $\lambda_T$.}
\label{sec:lambdat}
A priori, we do not know the magnitude of  $\sigma_T$.  However we estimate it
using the two gluon exchange model for the total cross section of R-hadron-nucleon collisions developed in ref. ~\cite{Gunion1} and used in ~\cite{Baer}.
This analysis suggests that the ratio
\begin{equation}
\frac{\lambda_T(R)}{\lambda_T(\pi)} \equiv  \frac{\sigma_T(\pi N)}{\sigma_T(R N)} =    (\frac{C_F}{C_A}) (\frac{<r^2_\pi>}{<r^2_R>})
\end{equation}
where $C_F = 4/3(C_A = 3)$ refers to the quadratic casimir of SU(3)$_{color}$ in the fundamental (adjoint) representations and $<r^2>$ is the transverse size-squared of the particle.

The first factor is due to the octet nature of $R$ constituents.
The reduced constituent mass of the particle determines $<r^2>$. In the case
of a pion we have $<r^2_\pi> \propto 4/m^2_q$ and in the case of $R^0$ we get
$<r^2_{R^0}>\propto 1/m^2_g$ (for $m_{\tilde{g}}\gg m_g$), where $m_q$ and $m_g$
are the light quark and gluon masses, respectively. BCG ~\cite{Baer} assume
equal values for $m_q$ and $m_g$ and estimate $\sigma^T_{R^0N}\sim (9/16)\sigma^T_{\pi N}$ which translates into $\lambda_T(R^0)\sim (16/9)\lambda_T(\pi)$.  Noting that $\lambda_T(\pi)\sim 11$ cm, they conclude that $\lambda_T(R^0)\sim 19$ cm is a reasonable estimate.  However the constituent quark mass is roughly speaking  $(1/3) m_{proton} \sim 330$ MeV and the constituent gluon mass is $(1/2) m_{glueball} \sim 750$ MeV.  Hence it might be more realistic to use the relation $m_g \sim 2m_q$ which changes the result to $\lambda_T(R^0)\sim 78$ cm.   In this analysis we take $\lambda_T=19$ cm and $\lambda_T=38$ cm, which are the values used by BCG ~\cite{Baer}.

It will be seen later that larger values of $\lambda_T(R^0)$ can open an allowed region in the gluino mass window.  Although we have adopted the estimates of BCG, we believe that choosing a larger value for $\lambda_T$ is NOT an extreme limit, but is in fact preferred.  We also assume that all R-hadrons have the same $\lambda_T$.   We rescale all the differential cross sections to get the desired $\lambda_T$ and keep their ratios unchanged.  Note, the ratios are determined by the Regge analysis discussed in the following section.

The average number of hadronic collisions inside the hadronic calorimeter depends on the magnitude of $\lambda_T$. For $\lambda_T=19$ $cm$, the R-hadrons on average undergo $6$ hadronic collisions in the hadronic calorimeter and $2$ hadronic collisions in the un-instrumented iron between the muon chambers.

\subsubsection{Regge scattering model: $\Delta E$ and charge fluctuation} 
\label{sec:regge}

The differential cross sections for the four R-hadrons determine both $\Delta E$, the energy loss per hadronic collision and the probability for charge fluctuation. In this analysis we use single particle inclusive Regge cross sections with Pomeron and $\rho$ Regge exchanges.  Recall that Regge exchange is
by far the dominant contribution to high energy hadronic scattering cross sections.

 Let us briefly review Regge theory. 
The Regge pole model connects the two classes of phenomena:

\begin{itemize}

\item classification of hadrons, and
  
\item high energy scattering of hadrons.

\end{itemize}

Hadronic bound states or resonances with identical 
quantum numbers, except spin and mass, are correlated by 
Regge trajectories where they appear with spin $J$ 
differing by two units. There is another relevant 
quantum number in Regge theory known as the signature
$\tau$. For mesons, signature is defined to be $\tau=(-1)^J$
with the corresponding particle sequences, $J=0,2,4,...,({\rm for} \;\; \tau=+1)$ and $J=1,3,5,...,( {\rm for} \;\; \tau=-1)$. As proposed by
Chew and Frautschi \cite{Chew},
particles are classified on Regge trajectories in plots 
of their spin $J$  versus the square of
their mass $M^2$. The trajectory $\alpha$, interpolates
between the particles such that $Re[\alpha(M^2_J)]=J$.\footnote{We 
only talk about meson trajectories here; baryon 
trajectories are similarly discussed in ref. \cite{Barger}.}  The
trajectory $\alpha(t)$ is assumed to be an analytic function of
$t$ and experimentally they are, to a good 
approximation, straight lines. The two body scatterings are dominated 
by the exchange of one or more of these trajectories. To see how a 
Regge pole exchange amplitude could arise in a field theory, 
consider the Van Hove-Durand model for the scattering of 
$A+B\to A+B$ with $A$ and $B$ spinless particles. $A$ and
$B$ interact via exchange of a meson with spin $J$ in the $t$-channel. 
The amplitude is given by
$$
A(s,t)=\frac{g^2_J[-s/s_0]^J}{M^2_J-t}
$$ 
in the large $s$ limit. If the exchanged particle is a member of
a Regge trajectory having an infinite series of mesons with spins
$J=1,3,5,...$, the scattering amplitude can be expressed as a summation
of the single exchanges of all the mesons on the trajectory. 
$$
A(s,t)=\sum_{J}\frac{g^2_J}{M^2_J-t}
[\frac{(-1)^J-1}{2}](\frac{s}{s_0})^J,
$$ 
where $s_0$ is an energy scale factor. Assuming a universal 
coupling and a linear Regge trajectory $M^2_J=\mu^2(J-a)$, the
series sum can be expressed in closed form. The result in the 
limit of large $s$ is
$$
A(s,t)=-\frac{g^2\pi}{2\mu^2}\frac{(1-e^{-i\pi\alpha})}{\sin(\pi\alpha)}
(\frac{s}{s_0})^\alpha,
$$
which is the form for the exchange of Regge poles with odd signature
with a trajectory $\alpha(t)=a+t/{\mu^2}$  satisfying $\alpha(M^2_J)=J$.
The trajectory $\alpha(t)$, for $t \leq 0$ describing scattering in the s channel is a smooth continuation of the Regge trajectory for
$t > 0$ describing scattering due to resonance exchange exchange in the t channel. In general, the form of a Regge amplitude can be more
complicated if the incoming particles carry spin or charge.

\subsubsection{INCLUSIVE SCATTERING, TRIPLE REGGEON FORM}
\label{sec:inclus}

\begin{figure}
\scalebox{1.0}[1.0]
{
\includegraphics{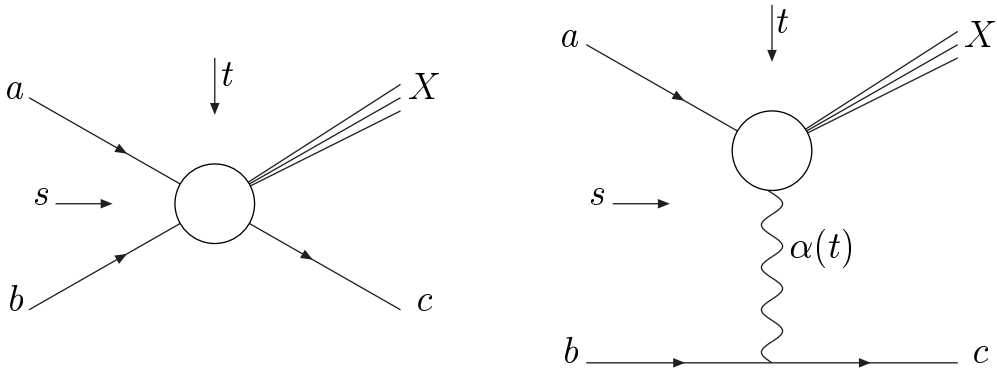}
}
\caption{Diagrams showing an inclusive
process $a+b\rightarrow c+X$ in terms of a Regge pole $\alpha(t)$ in the
t-channel}
\label{f:regge1}
\end{figure}

The single particle inclusive scattering process $a+b\rightarrow c+X$ can also be expressed in terms of Regge exchanges. $a$, $b$ and $c$ are
definite particles and $X$ represents everything else with total invariant mass 
$M_X$.  In the limit
\begin{eqnarray} 
M_X^2 \;\; {\rm large}, \;\; s/M_X^2 \gg 1 \;\; \rm and \;\; s\gg t  \label{eq:limit}
\end{eqnarray}
 the near forward two-body inclusive scattering process can be described by the t-channel exchange of a Reggeon.  Although
single particle inclusive scattering represents only $\sim 20$\% 
of the total hadronic cross section we will use this model for R-hadron scattering in our analysis.  It determines the energy loss $\Delta E$ as well as the probability for charge fluctuation.  This process is represented in Fig.\ \ref{f:regge1} . 

The amplitude for the process in Fig.\ \ref{f:regge1} is given by

\begin{eqnarray}
A(s,t) & \propto \beta_{b\bar{c}}(t) \, \xi(t) \, s^{\alpha(t)}. &
\end{eqnarray}
   
\begin{figure}
\scalebox{1.0}[1.0]
{
\includegraphics{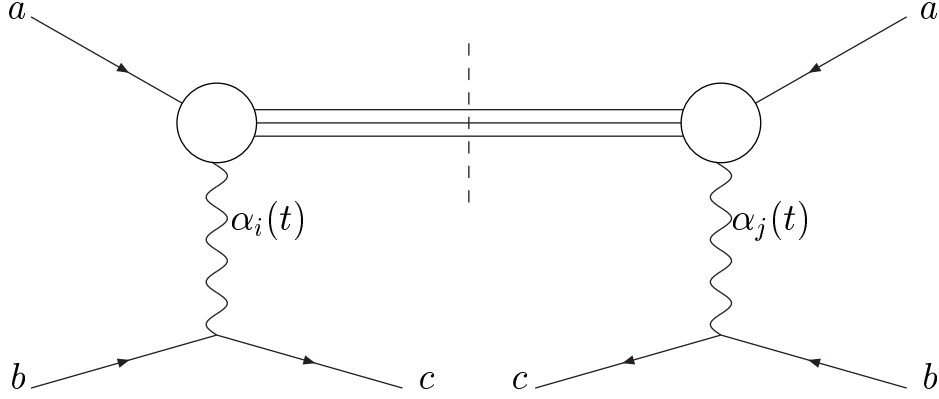}
}
\caption{Using a generalized optical theorem due to Mueller \cite{Mueller} we
relate the single particle inclusive cross section $a+b\rightarrow c+X$ to the discontinuity in the forward three-body
amplitude $a+b+\bar{c}\rightarrow a+b+\bar{c}$}.
\label{f:regge2}
\end{figure}

\begin{figure}
\scalebox{1.0}[1.0]
{
\includegraphics{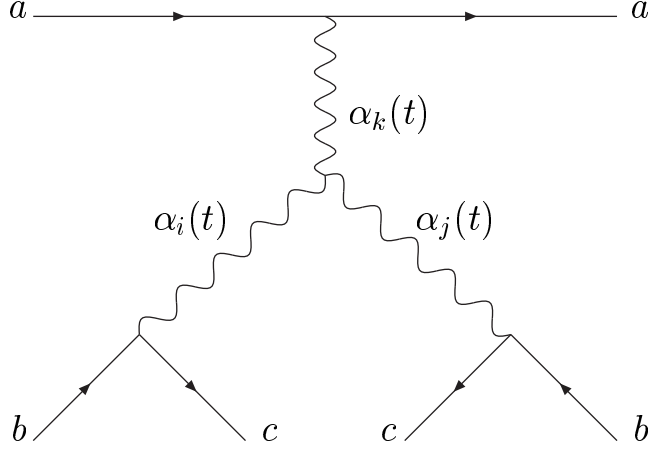}
}
\caption{The triple Reggeon scattering diagram describing
the diffractive process $a+b\rightarrow c+X$}
\label{f:regge3}
\end{figure}

Using a generalized optical theorem due to Mueller \cite{Mueller} as shown
in Fig.\ \ref{f:regge2} the inclusive
cross section $a+b\rightarrow c+X$ is
related to the discontinuity in the forward three-body
amplitude $a+b+\bar{c}\rightarrow a+b+\bar{c}$.  In the Regge limit (eqn. ~\ref{eq:limit}) this gives the triple Regge cross section (see Fig. \ref{f:regge3})

\begin{equation}
\label{eq:1}
\frac{d^2\sigma}{dxdt}=\frac{1}{s}\sum_{i,j,k}
G_{ijk}(t)(\frac{s}{M_X^2})^{\alpha_i(t)+\alpha_j(t)}(M_X^2)^{\alpha_k(0)} ,
\end{equation}

\noindent with $x=1-\frac{M_X^2}{s}$.

For simplicity we consider two Regge trajectories, an
isosinglet Pomeron trajectory $P$ and an isovector Reggeon trajectory $\rho$. We
then write all the possible triple couplings of the Pomeron, Reggeon
and R-hadrons as follows

\begin{equation}
\label{eq:2}
g_{RRP} \, R^1_0R^2_0P +
g_{\tilde{\rho}\tilde{\rho}P} \, \tilde{\rho}^1_i\tilde{\rho}^2_iP +
g_{R\tilde{\rho}\rho} \, R_0\tilde{\rho}_i\rho_i +
g_{\tilde{\rho}\tilde{\rho}\rho} \,
\epsilon_{ijk}\tilde{\rho}^1_i\tilde{\rho}^2_j\rho_k,
\end{equation}
where the indices 1 and 2 distinguish between the outgoing and incoming 
particles in a vertex.

Using
\begin{equation}
\label{eq:3}
\rho_{\pm}=\frac{\rho_1\pm i\rho_2}{\sqrt{2}},
\quad
\tilde{\rho}_{\pm}=\frac{\tilde{\rho}_1 \pm i\tilde{\rho}_2}{\sqrt{2}}.
\end{equation}
 we find
\begin{eqnarray}
\tilde{\rho}_i\rho_i =& \tilde{\rho}_+\rho_-+\tilde{\rho}_-\rho_+
+\tilde{\rho}_3\rho_3, &  \\
\tilde{\rho}^1_i \tilde{\rho}^2_i =& \tilde{\rho}^1_+ \tilde{\rho}^2_- +
\tilde{\rho}^1_- \tilde{\rho}^2_+ + \tilde{\rho}^1_3 \tilde{\rho}^2_3, & \nonumber \\
\epsilon_{ijk}\tilde{\rho}^1_i\tilde{\rho}^2_j\rho_k & =
i(+\tilde{\rho}^1_+\tilde{\rho}^2_-\rho_3-
\tilde{\rho}^1_-\tilde{\rho}^2_+\rho_3+
\tilde{\rho}^1_-\tilde{\rho}^2_3\rho_+ & \nonumber \\
& - \tilde{\rho}^1_+\tilde{\rho}^2_3\rho_--
\tilde{\rho}^1_3\tilde{\rho}^2_-\rho_++
\tilde{\rho}^1_3\tilde{\rho}^2_+\rho_-). & \nonumber 
\end{eqnarray}

We thus obtain the following vertices and the relevant couplings given by
\begin{eqnarray}
R_0 R_0 P  &\rm with & g_{RRP}, \\
\tilde{\rho}_+ \tilde{\rho}_- P, \;
\tilde{\rho}_- \tilde{\rho}_+ P \;\; \rm and \;\;
\tilde{\rho}_3 \tilde{\rho}_3P & \rm with & 
g_{\tilde{\rho}\tilde{\rho}P},  \nonumber \\
R_0\tilde{\rho}_-\rho_+, \;
R_0\tilde{\rho}_+\rho_- \;\; \rm and \;\;
R_0\tilde{\rho}_3\rho_3 & \rm with &
g_{R\tilde{\rho}\rho} , \nonumber \\
+\tilde{\rho}_+ \tilde{\rho}_- \rho_3,\; 
-\tilde{\rho}_- \tilde{\rho}_+ \rho_3,\; 
+\tilde{\rho}_- \tilde{\rho}_3 \rho_+, &  & \nonumber \\ 
-\tilde{\rho}_+ \tilde{\rho}_3 \rho_-, \; 
-\tilde{\rho}_3 \tilde{\rho}_+ \rho_- \;\; \rm and \;\;
+\tilde{\rho}_3 \tilde{\rho}_- \rho_+ &\rm with & \;\;
ig_{\tilde{\rho}\tilde{\rho}\rho}.  \nonumber
\end{eqnarray}
which lead to the following diffractive scattering patterns,
$$
R_0 N {\to} R_0 X \Longrightarrow G^1_{PPP},
$$ 
$$
R_0 N {\to} \tilde{\rho}_{(+-3)} X, \;\;
\tilde{\rho}_{(+-3)} N {\to} R_0 X 
\Longrightarrow G^1_{\rho\rho P},
$$ 
$$
\tilde{\rho}_{+} N {\to} \tilde{\rho}_{+} X, \;\;
\tilde{\rho}_{-} N {\to} \tilde{\rho}_{-} X
\Longrightarrow G^2_{PPP},G^2_{\rho\rho P},
$$
$$
\tilde{\rho}_{3} N {\to} \tilde{\rho}_{3} X   
\Longrightarrow G^2_{PPP},
$$
$$
\tilde{\rho}_{+} N {\to} \tilde{\rho}_{-} X, \;\;
\tilde{\rho}_{-} N {\to} \tilde{\rho}_{+} X
\equiv 0,
$$
$$
\tilde{\rho}_{+} N {\to} \tilde{\rho}_{3} X, \;\;
\tilde{\rho}_{3} N {\to} \tilde{\rho}_{+} X, \;\;
\tilde{\rho}_{-} N {\to} \tilde{\rho}_{3} X, \;\;
\tilde{\rho}_{3} N {\to} \tilde{\rho}_{-} X 
\Longrightarrow G^2_{\rho\rho P}.
$$

Using eq. \ref{eq:1} and the Pomeron and Reggeon trajectories
\begin{equation}
\label{eq:4}
\alpha_P(t)=1+\breve{\alpha}_P t,\quad
\alpha_\rho(t)=\alpha+\beta t,
\end{equation}
we obtain the desired differential scattering cross section 
formulas for the triple Pomeron contribution,
\begin{eqnarray}
\frac{d^2\sigma_{PPP}}{dM_Xdt} & = & 2G_{PPP}(t)
(\frac{s}{M_X^2})^{2\breve{\alpha}_P t}(\frac{1}{M_X}),
\end{eqnarray}
and for the Reggeon-Reggeon-Pomeron contribution,
\begin{eqnarray}
\frac{d^2\sigma_{\rho\rho P}}{dM_Xdt} & = & 2G_{\rho\rho P}(t)
(\frac{s}{M_X^2})^{(2\alpha+2\beta t-2)}(\frac{1}{M_X}).
\end{eqnarray}
These results are used to perform the analysis in this paper.\footnote{For simplicity we have ignored the triple Reggeon contribution.   We also use the Regge cross-sections for all physical values of s, t and $M_X$; in particular, extending outside the Regge limit (eqn. 2).} The calculations are based on the assumption that
\begin{eqnarray}
G_{PPP}(t)=G^1_{PPP}(t)=G^2_{PPP}(t), \;\;&
G_{\rho\rho P}(t)=G^1_{\rho\rho P}(t)=G^2_{\rho\rho P}(t).&
\end{eqnarray}
Inspired by ref. \cite{Field}, we take 
\begin{eqnarray}
G_{PPP}(t)/ G_{PPP}(0) =  & 0.88 \ e^{3.94 \ t} \, + \, 0.12 \ e^{1.12 \ t} & \\
G_{\rho\rho P}(t)/ G_{\rho\rho P}(0) = &  0.85 \ e^{7.26 \ t} \, + \, 0.15 \ e^{-1.83 \ t}. & \nonumber
\end{eqnarray} and
\begin{eqnarray}
\breve{\alpha}_P = & 0.36 &  \\
\alpha = &  0.5 &  \nonumber \\
\beta  = &  1.0  &  . \nonumber 
\end{eqnarray}

We also vary
the relative size of $G_{PPP}(t)$ and $G_{\rho\rho P}(t)$ to check the
sensitivity of our results to these couplings.

Before continuing, let us briefly discuss the scattering analysis of BCG \cite{Baer}.   These authors assume that when an R-hadron scatters on a nucleon the light brown muck (quarks and gluons bound to the gluino) are stripped off.  The bare gluino then re-fragments into a {\em charged} R-hadron with the fragmentation probability $P$ and into a {\em neutral} R-hadron with probability $1 - P$.   Thus for BCG, the probability $P$ plays two independent roles:
\begin{itemize}
\item fragmentation probability, and
\item  rescattering probability.
\end{itemize}

In our analysis, however,  Regge scattering cross sections allow R-hadrons to change their identity with velocity dependent probabilities in each hadronic collision.  We thus separate the independent phenomena of fragmentation and rescattering. 

In addition, to calculate the energy of an R-hadron after an R-hadron - nucleon collision, BCG ~\cite{Baer} use either a constant differential cross section which vanishes for $t>1$ $GeV$ or a triple Pomeron scattering formula.   We on the other hand use the differential Regge cross sections for this as well.

\section{THE ANALYSIS}
\label{sec:how}

Consider the CDF data for jets+${\not p}_T$ \cite{Exper}.   We compare our Monte Carlo simulation with this data and thus we use the same cuts as BCG, which were designed to duplicate the experimental procedures of ref. \cite{Exper}. The cuts are listed as follows.

\begin{enumerate}
\Roman{enumiv}
\item No (isolated) leptons with $E_T>10$ GeV.
  
\item ${\not p}_T>60$ GeV.

\item There are three or more jets with $|\eta_{jet}|<2$ 
and $E_T>15$ GeV, using a coalescence cone size of $\Delta R=0.5.$
                   
\item Azimuthal separation requirements: $\Delta\phi({\not p}_T,j_1)<160^\circ$
and\\ $\Delta\phi({\not p}_T,j(E_T>20GeV)) > 30^\circ$.  

\end{enumerate}

For our monte carlo analysis we generated events using ({\it SPYTHIA, A
Supersymmetric Extension of PYTHIA 5.7} \cite{spythia}), which has been modified by S. Mrenna and K. Tobe to accommodate a gluino LSP.  We have written a toy
calorimeter code extending out to $|\eta_{jet}|<4$. The cell size
of the calorimeter is set to $\Delta\eta\times\Delta\phi=0.1\times0.1$. The
hadronic resolution is also taken to be $70\%/\sqrt{E}$. The missing
transverse momentum cut (2) and the jet-number cut (3) are the strongest
cuts. $j1$ is the most energetic jet in an event and $j(E_T>20GeV)$ is 
any jet with a transverse energy larger than $20$ GeV.   Cut (4) eliminates
the QCD jet mis-measurement backgrounds. These backgrounds occur due to the uncertainty in the energy measurement of very energetic jets leading to large missing momentum in either the same or opposite direction as their momenta.

\subsection{Identifying missing momentum}
\label{sec:mom}

 From the total energy deposited in the hadronic calorimeter and the pseudo-rapidity $\eta$ of the corresponding jet, one can determine the measured transverse energy of the jet.  We assume (like CDF) that the jets are massless.  Hence the magnitude of the transverse momentum is the same as the transverse energy.

If we add up the transverse momenta of all jets present in a single detected event,  we should get zero due to the conservation of transverse momentum. In practice this does not happen.  For one reason, not all the generated particles are energetic enough to be detected by a calorimeter cell as part of a 
jet (this effect is parameterized in pythia).   The second and very important reason is that an R-hadron does not deposit all its energy inside the calorimeter; resulting in some missing momentum.  For example, (1) only the
kinetic energy of an R-hadron is visible in a detector and not its rest 
mass and (2) the energy loss per collision decreases as the R-hadron mass increases.  There is also an enhancement in the ionization energy loss for R-hadrons considered later in Sec. \ref{sec:ion}.  The total missing transverse momentum ${\not p}_T$ is defined as minus the sum of the transverse momenta of all jets in each event. A large missing transverse momentum is a signature for a gluino LSP.  

Spythia gives us information which is directly converted into $p_T$ for all standard model particles.  For R-hadrons on the other hand the output of Spythia is input for our hadronic calorimeter code.  It is in this code that $\lambda_T$ and Regge cross sections are used to determine the energy deposited in the hadronic calorimeter.  As discussed earlier, R-hadrons suffer both hadronic and ionization energy loss in the hadronic calorimeter.  Hadronic energy losses are due to collisions of R-hadrons with nuclei.  We assume an R-hadron - nucleon collision occurs in the hadronic calorimeter after each distance $\lambda_T$ in the detector.  We use the single particle inclusive differential Regge cross sections, discussed in sec. ~\ref{sec:inclus},
to obtain the energy and charge of the R-hadrons after each hadronic collision.  These are computed randomly on an event by event basis using a probability distribution which is weighted by the differential cross sections.  Each R-hadron can scatter into any of the four R-hadrons after colliding with a nucleon.  Therefore there are four separate processes which can occur.  To decide which one occurs, we compute the total cross section for each process. The ratio of the total cross sections determines the probability of each process.  A random number is generated which decides which process occurs based on these probabilities.   In the next step we use the corresponding differential cross section to determine the energy of the out-going R-hadron.  The energy of an out-going R-hadron is a function of the momentum transfer $t$ and the invariant mass $M_X$ of the inclusive process.  The probability density for an R-hadron with given $t$, $M_X$ and center of momentum energy $\sqrt{s}$
is  $ \frac{d^2 \sigma}{dt dM_X}|_{t, M_X}$.  A ($t, M_X)$ pair is therefore generated, weighted by $ \frac{d^2 \sigma}{dt dM_X}|_{t, M_X}$ to compute the energy of the out-going R-hadron.

\subsection{Ionization energy loss}
\label{sec:ion}
To calculate the ionization energy deposit, we use the standard
$dE/dx$ formula from ref. ~\cite{Data}.  A $50$ $GeV$ 
muon beam, as measured in the central tracking chambers by CDF, deposits $2$ $GeV$ of energy in the hadronic calorimeter. If we compute the energy loss of the muon beam using the standard $dE/dx$ formula for a muon passing through an 
iron detector, we find it deposits only $1.3$ $GeV$.  Ref. \cite{Baer} thus defines the ratio of the measured calorimeter ionization loss to the actual $dE/dx$ loss from ionization, $E_{ionization}$, as $r = 1.6$.   Finally, the visible energy of a jet (that measured in the calorimeter) is the energy we assign to the jets when calculating  ${\not p}_T$.  It is given by the total energy loss $E_{total}=rE_{ionization}+E_{hadronic}$.  As mentioned earlier all jets are massless and we identify the magnitude of the 3-momentum of a jet with $E_{total}$ and choose its original direction as the direction of its momentum 3-vector. In this way, we can add up the 3-vector momenta of all jets present in a single event and determine the total ${\not p}_T$.  
      
\subsection{R-hadrons identified as muons}
{\em Some of the R-hadrons which pass through the CDF detector are identified as muons.}  An R-hadron jet is declared to be muon-like, if it is
detected to be charged in the central tracker and in at least one of the muon 
chambers, covering only a portion of the available pseudo-rapidity region. In addition, it should have a momentum larger than $10$ $GeV$ in the 
central tracker and also should not deposit more than $6$ $GeV$ inside the hadronic calorimeter as measured by the detector.  Each event containing a muon-like jet is discarded~\cite{Exper}.  In some cases a large number of events are discarded because they contain muon-like jets.

It is thus important to understand under what circumstances an R-hadron jet will be muon-like.  Neutral R-hadron jets will not be muon-like since the jets need to be charged both in the central tracker and in one of the muon chambers.  On the other hand, if a jet remains charged for most of its passage through the hadronic calorimeter, it is likely that it will deposit too much energy (more than $6$ $GeV$) and thus will not be a muon candidate.  Hence an ideal muon-like jet is one that is charged in the central tracker and one of the 
muon chambers and is neutral for most of its passage through the
hadronic calorimeter. From the above discussion, we understand that increasing
$P_{\rho^+}$ and $P_{\rho^-}$ will increase the number of R-hadron jets that are
identified as muon jets.   However if $\sigma_{PPP} \gg \sigma_{\rho \rho P}$ (i.e. Pomeron scattering dominates) then there is no chance for charge fluctuations and typically the charged R-hadron deposits too much energy in the hadronic calorimeter.  Finally, if $\lambda_T $ increases then there are fewer hadronic collisions and thus less hadronic energy loss in the calorimeter.  In addition, as a consequence of fewer hadronic collisions, the R-hadron retains its velocity.  This results in less ionization energy loss since a fast moving charged particle loses less energy due to ionization than a slower particle, if its velocity is below minimum ionizing which is the case most of the time here.   This tends to increase the number of muon-like events.

\section{THE RESULTS}
\label{sec:result}

To determine the limits on the mass of the gluino LSP, we compare the cross
section for gluino pair production, after applying the cuts, with the
standard model background as reported in the analysis by CDF ~\cite{Exper}.  They report a background rate of $28.7$ events~\footnote{This background rate is determined using a Monte Carlo calculation.  The measured background is actually 36 events.} for an integrated luminosity of $19$ $pb^{-1}$ at the Fermilab center of mass energy of $\sqrt{s}=1.8$ $TeV$ corresponding to $\sigma_B=1.51$ $pb$.  A $1.96\sigma$ background fluctuation corresponds to an allowed signal of $\sigma_s\sim 553$ $fb$, which is also equivalent to $\sigma_s\sim 614$ $fb$ with an integrated luminosity of $0.1$ $fb^{-1}$ at $5\sigma$ level.  This is the bound used by BCG~\cite{Baer}. We also use the latter cross section and do our analysis for an integrated luminosity of $0.1$ $fb^{-1}$.   

If the gluino signal, after cuts, is larger than $\sigma_s\sim 614$ $fb$ with an integrated luminosity of $0.1$ $fb^{-1}$ then those specific values of the gluino mass, $\lambda_T$, fragmentation probabilities and Reggeon couplings, 
are excluded.   We first analyze this parameter space in the limit that all other superparticles are too heavy to be produced at the Tevatron.  We then determine the sensitivity of our results  to lighter squarks by lowering the squark masses. $\lambda_T$, $\sigma_{PPP}/\sigma_{\rho\rho P}$ and the {\it fragmentation probability vector}  $P=(P_{R^0},P_{\rho^+},P_{\rho^0},P_{\rho^-})$ are also varied to check the sensitivity of our results to these parameters.  The labels assigned to the different parameter choices discussed below are given in table \ref{t:parameters}.

\protect
\begin{table}
\caption[8]{
{\it Labels for the different parameter choices used in the figures with $\lambda_T$ (interaction length),  the ratios of triple Regge vertices and
the fragmentation probability vector $P = (P_{R^0}, P_{\rho^+}, P_{\rho^0}, P_{\rho^-})$. } \\
% end of caption
}
\label{t:parameters}
$$
\begin{array}{|l|l|}
\hline
{\rm Labels} & {\rm Parameter \;\; values}  \\
\hline
l1, \;\; l2 &  \lambda_T = 19 cm, \;\; \lambda_T = 38 cm \\
\hline 
g1, \;\; g2 &  \sigma_{PPP} \sim \sigma_{\rho\rho P}, \;\;  \sigma_{PPP}\gg \sigma_{\rho\rho P}   \\
\hline 
p1, \;\; p2, \;\; p3 &  
P=(1/4,1/4,1/4,1/4), \;\; P=(0,1,0,0), \;\; P=(1,0,0,0)
   \\
\hline 
\end{array}
$$
\end{table}

\begin{figure}
\scalebox{0.8}[0.8]
{
\includegraphics{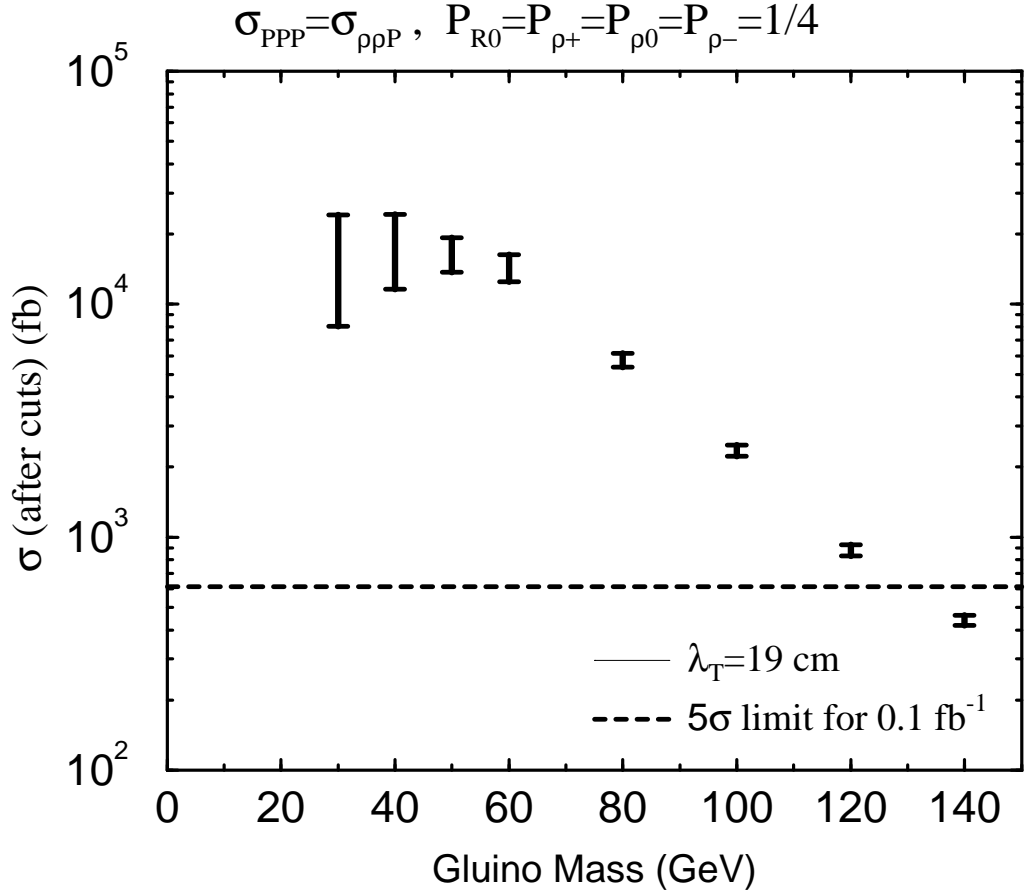}
}
\caption{The cross sections after cuts in the jets+${\not p}_T$ channel is compared to the 5$\sigma$ level for $L=0.1 \; fb^{-1}$ which is roughly the same as the $95\%$ CL for $L=19 \; pb^{-1}$ at $\sqrt{s}=1.8$ TeV. $\lambda_T=19$ cm and the Pomeron contribution is comparable with the Reggeon contribution to the total scattering cross section. The gluino has initially fragmented into the neutral and charged R-hadrons with equal probabilities.}
\label{f:equal19}
\end{figure}

Fig.\ \ref{f:equal19} corresponds 
to $\lambda_T=19$ cm, $\sigma_{PPP}\sim \sigma_{\rho\rho P}$ and
$P_{R^0}=P_{\rho^+}=P_{\rho^0}=P_{\rho^-}=1/4$. Using our labeling system,
this is called a $(l1g1p1)$ scenario. As seen in this case, we are able to exclude gluino masses from $30$ GeV up to $130$ GeV
at $95\%$ confidence level.  The large signal cross section at $30$ GeV suggests that the gluino mass exclusion can be pushed to even lower values.  Considering the fact that BCG have excluded gluinos with mass in the range $3$ to $22-25$ GeV using LEP data,  the gluino window is closed from $3$ to $130$ GeV for these scattering parameters.  It will become evident that the most important parameter is $\lambda_T$.   

\begin{figure}
\scalebox{0.8}[0.8]
{
\includegraphics{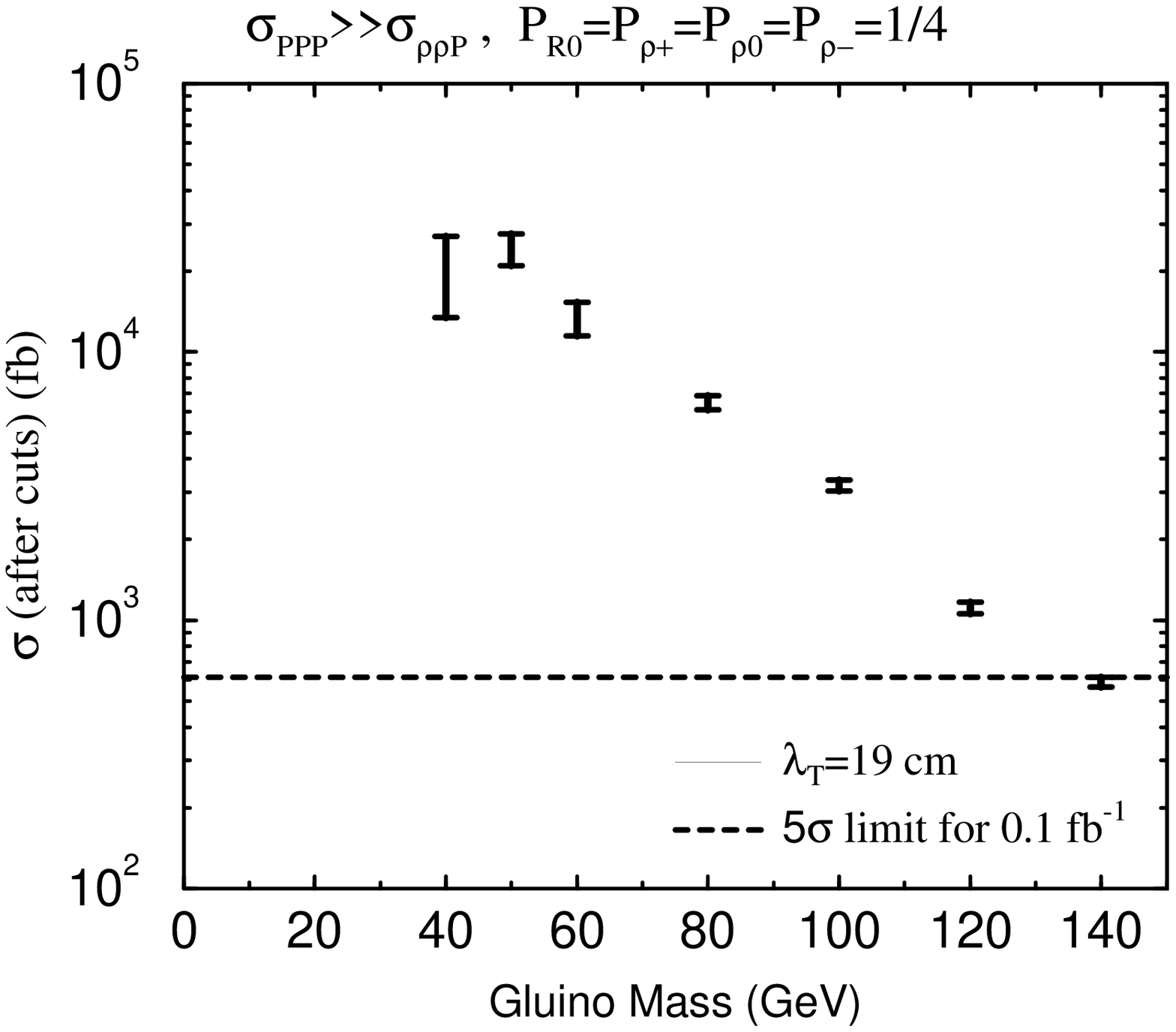}
}
\caption{The cross sections after cuts in the jets+${\not p}_T$ channel is compared to the 5$\sigma$ level for $L=0.1 \; fb^{-1}$ which is roughly the same as the $95\%$ CL for $L=19 \; pb^{-1}$ at $\sqrt{s}=1.8$ TeV. $\lambda_T=19$ cm and the Pomeron contribution is much greater than the Reggeon contribution to the total scattering cross section. The gluino has initially fragmented into the neutral and charged R-hadrons with equal probabilities.}
\label{f:10019}
\end{figure}

Fig.\ \ref{f:10019} corresponds to $(l1g2p1)$. In this case an R-hadron which
scatters off the nucleon will almost always retain its identity. The charged R-hadrons, produced $50\%$ of the time in this case, will always remain charged as they pass through the calorimeter. In $(l1g1p1)$, it is possible for a gluino to fragment into a charged R-hadron but become neutral in the muon chambers. This will not happen for the case of $(l1g2p1)$. We might expect an increase in the number of muon-like jets in $(l1g2p1)$ as compared to $(l1g1p1)$. On the other hand, there is a large ionization energy deposit for the charged R-hadrons in $(l1g2p1)$ because they remain charged all the way through the calorimeter. This can suppress the number of muon-like jets due to the fact that it would be less likely for them to deposit less that $6$ GeV 
of transverse energy in the calorimeter. The second effect dominates over the first and we see a decrease in the total number of muon-like jets in $(l1g2p1)$ compared to $(l1g1p1)$.  Looking at Fig.\ \ref{f:equal19} and Fig.\ \ref{f:10019} we can see this effect which shows itself as an increase in the signal cross section, mostly at large $m_{\tilde g}$, in Fig.\ \ref{f:10019}. In $(l1g2p1)$ the gluino mass is excluded up to $140$ GeV.

\begin{figure}
\scalebox{0.8}[0.8]
{
\includegraphics{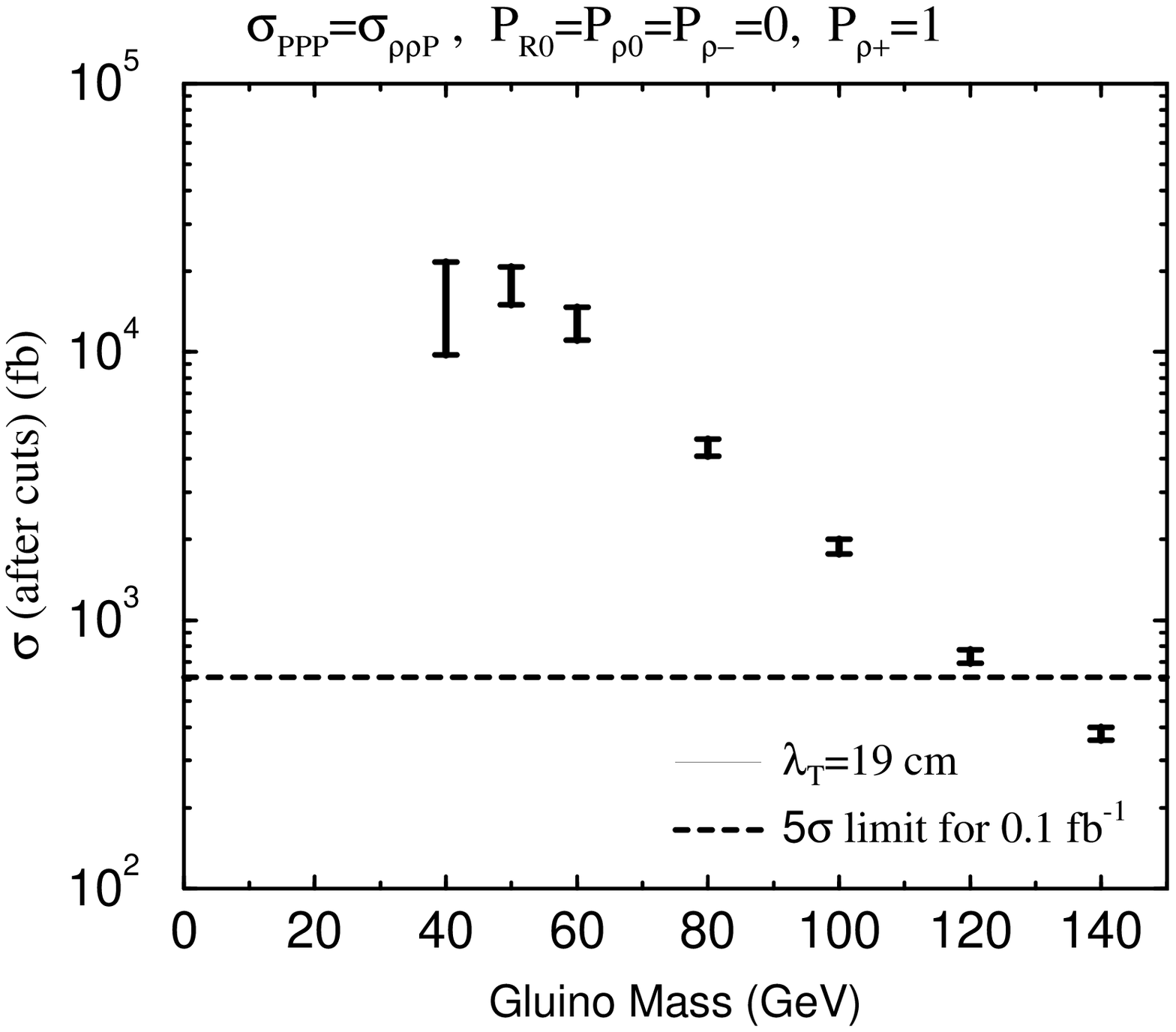}
}
\caption{The cross sections after cuts in the jets+${\not p}_T$ channel is compared to the 5$\sigma$ level for $L=0.1 \; fb^{-1}$ which is roughly the same as the $95\%$ CL for $L=19 \; pb^{-1}$ at $\sqrt{s}=1.8$ TeV. $\lambda_T=19$ cm and the Pomeron contribution is comparable with the Reggeon contribution to the total scattering cross section. The gluino has initially fragmented only into the charged R-hadrons.}
\label{f:eqr319}
\end{figure}

\begin{figure}
\scalebox{0.8}[0.8]
{
\includegraphics{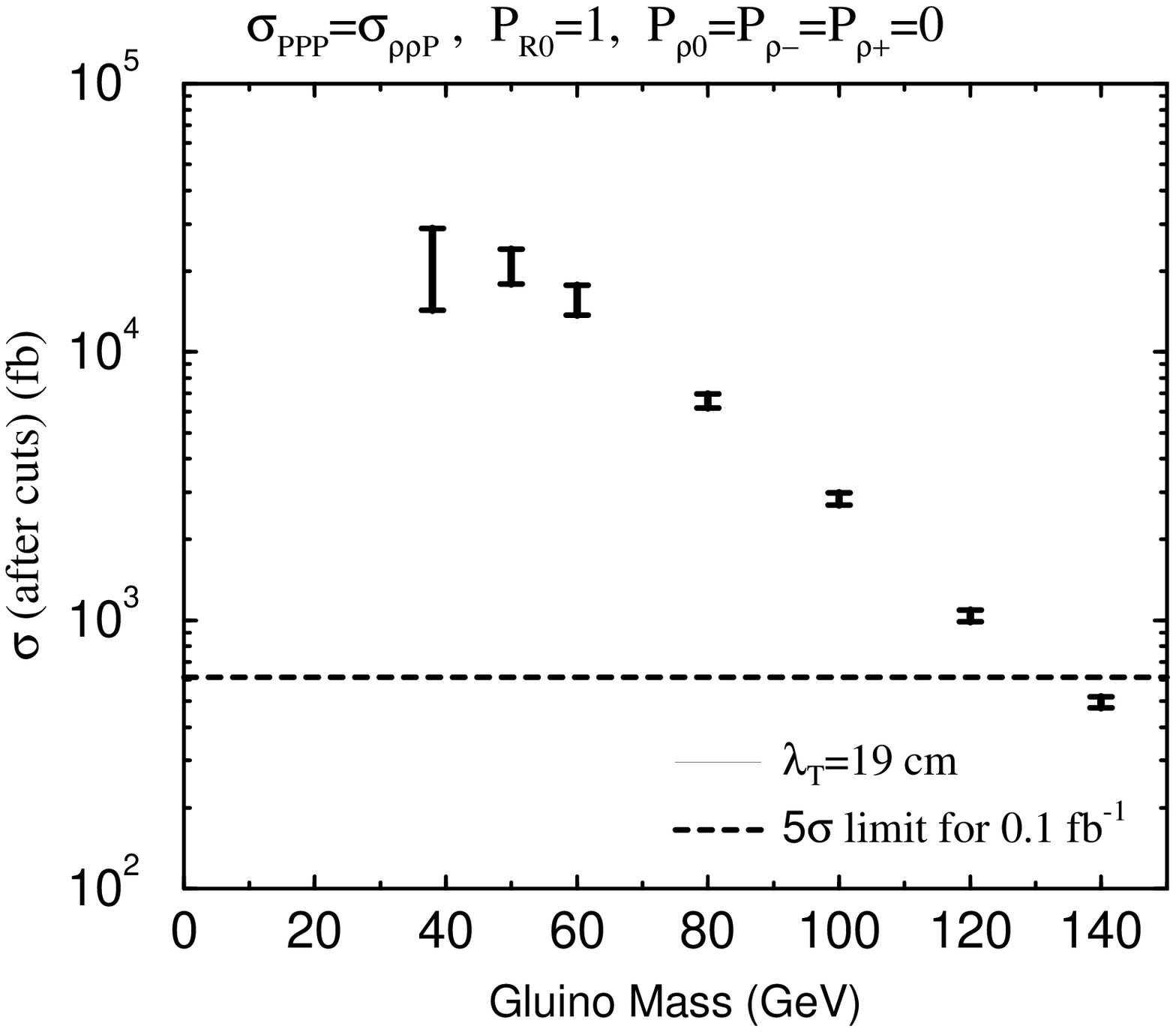}
}
\caption{The cross sections after cuts in the jets+${\not p}_T$ channel is compared to the 5$\sigma$ level for $L=0.1 \; fb^{-1}$ which is roughly the same as the $95\%$ CL for $L=19 \; pb^{-1}$ at $\sqrt{s}=1.8$ TeV. $\lambda_T=19$ cm and the Pomeron contribution is comparable with the Reggeon contribution to the total scattering cross section. The gluino has initially fragmented only into the neutral R-hadrons.}
\label{f:eqr019}
\end{figure}

\begin{figure}
\scalebox{0.8}[0.8]
{
\includegraphics{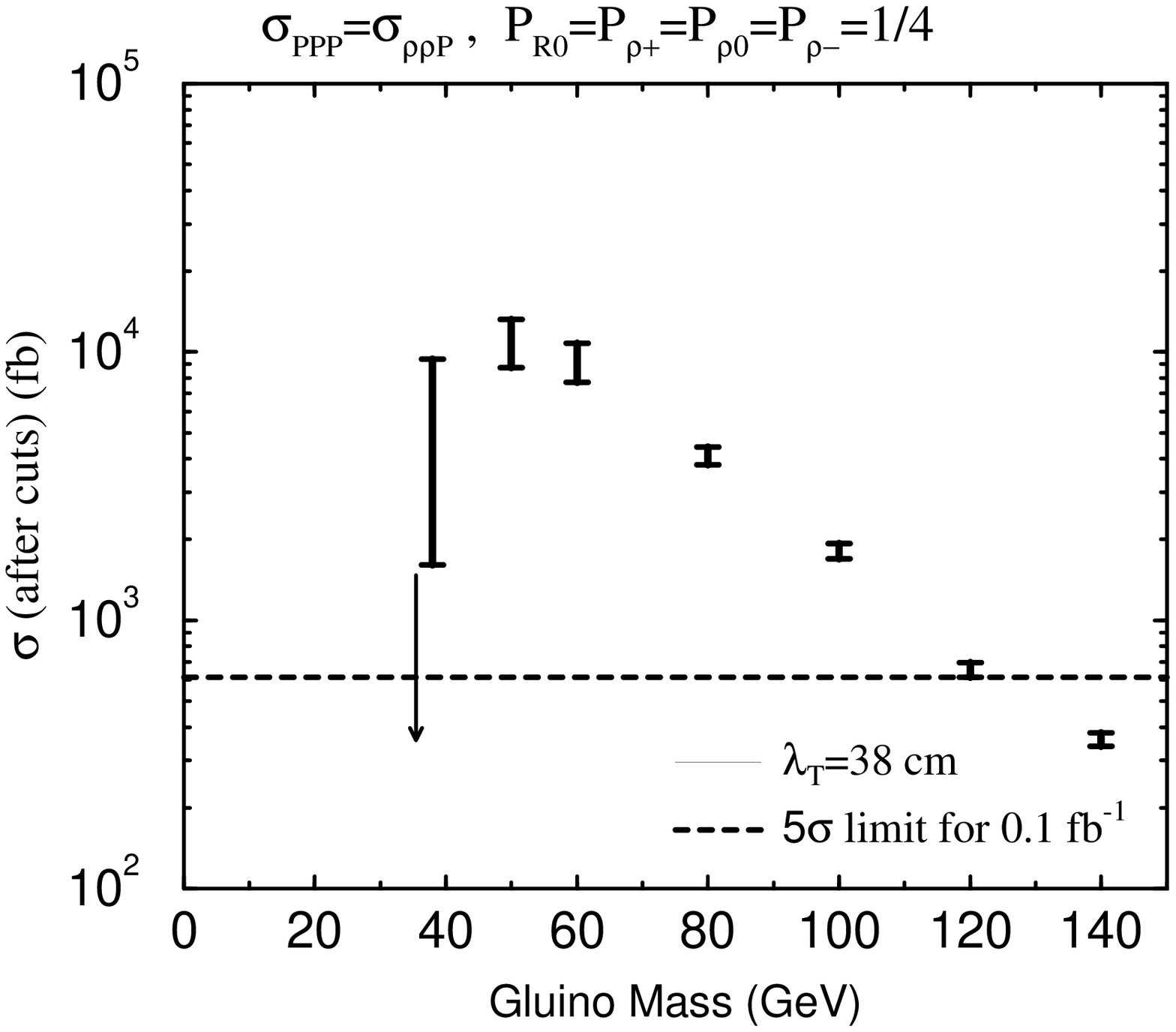}
}
\caption{The cross sections after cuts in the jets+${\not p}_T$ channel is compared to the 5$\sigma$ level for $L=0.1 \; fb^{-1}$ which is roughly the same as the $95\%$ CL for $L=19 \; pb^{-1}$ at $\sqrt{s}=1.8$ TeV. $\lambda_T=38$ cm and the Pomeron contribution is comparable with the Reggeon contribution to the total scattering cross section. The gluino has initially fragmented into the neutral and charged R-hadrons with equal probabilities.}
\label{f:equal38}
\end{figure}

\begin{figure}
\scalebox{0.8}[0.8]
{
\includegraphics{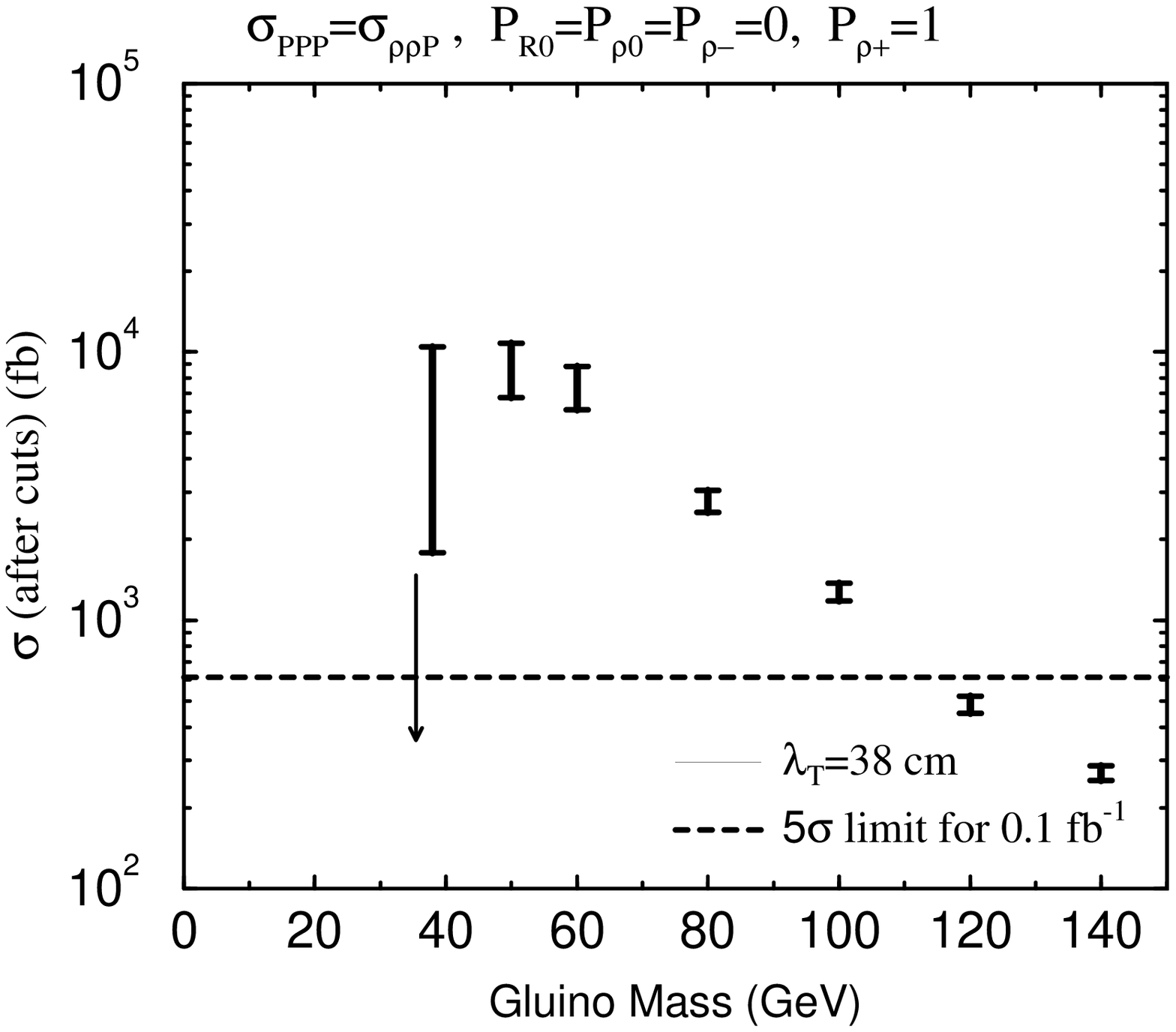}
}
\caption{The cross sections after cuts in the jets+${\not p}_T$ channel is compared to the 5$\sigma$ level for $L=0.1 \; fb^{-1}$ which is roughly the same as the $95\%$ CL for $L=19 \; pb^{-1}$ at $\sqrt{s}=1.8$ TeV. $\lambda_T=38$ cm and the Pomeron contribution is comparable with the Reggeon contribution to the total scattering cross section. The gluino has initially fragmented only into the charged R-hadrons.}
\label{f:eqr338}
\end{figure}

\begin{figure}
\scalebox{0.8}[0.8]
{
\includegraphics{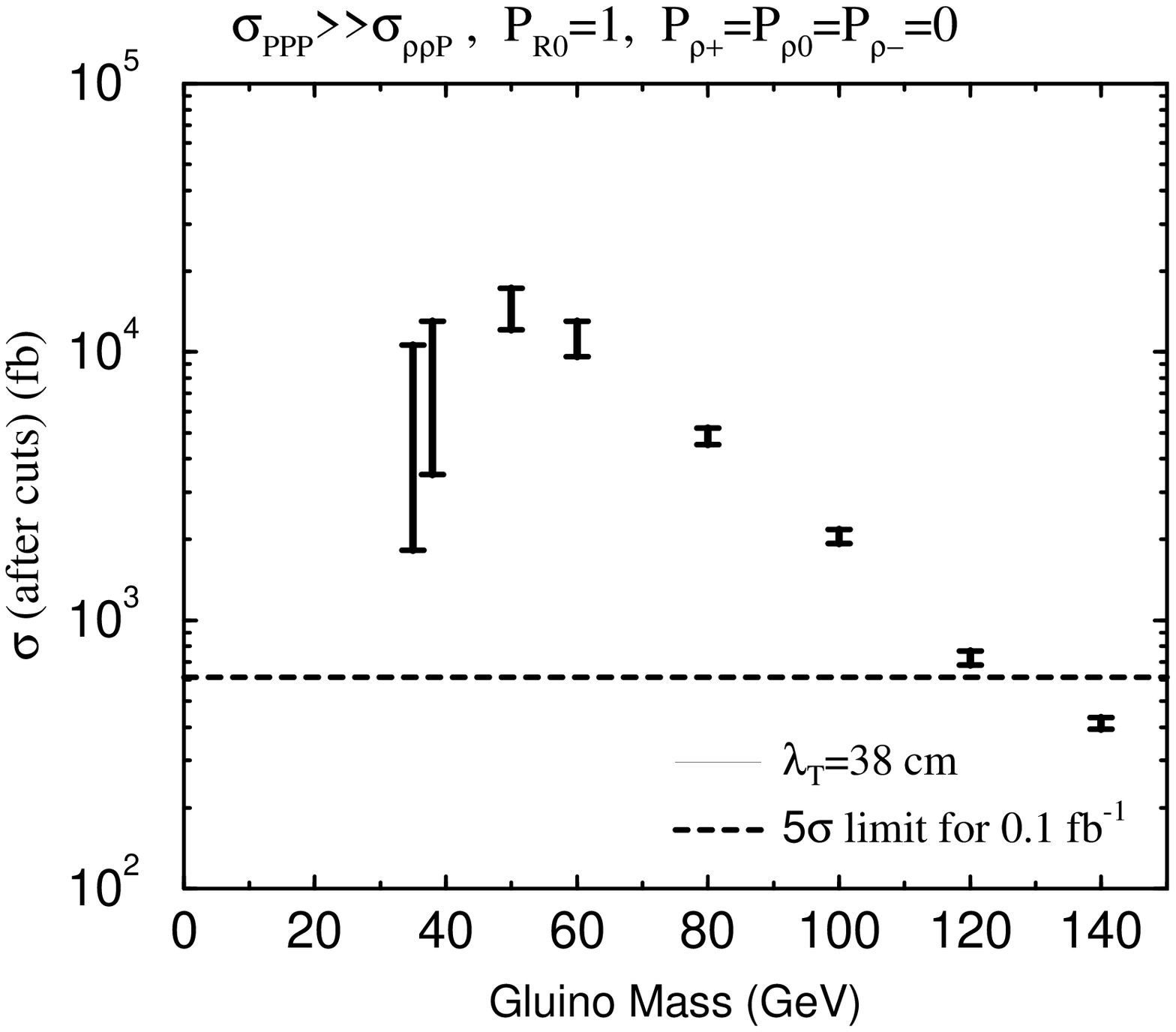}
}
\caption{The cross sections after cuts in the jets+${\not p}_T$ channel is compared to the 5$\sigma$ level for $L=0.1 \; fb^{-1}$ which is roughly the same as the $95\%$ CL for $L=19 \; pb^{-1}$ at $\sqrt{s}=1.8$ TeV. $\lambda_T=38$ cm and the Pomeron contribution is much greater than the Reggeon contribution to the total scattering cross section. The gluino has initially fragmented only into the neutral R-hadrons.}
\label{f:100r038}
\end{figure}

Fig.\ \ref{f:eqr319} is labeled ($l1g1p2$) where, in contrast to ($l1g1p1$), all gluinos initially fragment into charged R-hadrons. This greatly increases the chance of detecting a muon-like jet and results in a smaller signal cross section particularly at larger masses.  At smaller gluino mass, the missing momentum signal is reduced.  This is a consequence of two effects; for lighter gluinos (1) the hadronic energy loss per collision is larger and (2) the unmeasured mass energy is smaller.  Hence only a small number of events survive to be analyzed for the muon cut at small masses and the difference due to the muon jet cut is not significant between ($l1g1p2$) and ($l1g1p1$) at small gluino mass. Overall, we see a smaller signal cross section in Fig.\ \ref{f:eqr319} compared to ($l1g1p1$) in Fig.\ \ref{f:equal19}. The result is that we only exclude gluino masses up to $ \sim 125$ GeV.

($l1g1p3$) is opposite to ($l1g1p2$) where now all gluinos initially fragment into neutral R-hadrons.  Hence there are no muon-like jets present in this scenario. We expect larger cross sections as seen in Fig.\ \ref{f:eqr019}.  However as in ($l1g1p2$), we do not see a significant increase in the cross section at small gluino mass. We now exclude gluino masses up to $\sim 135$ GeV.  

In all the above cases, the signal cross section at small gluino mass is similar for the reason explained previously.   We therefore find, for $\lambda_T=19$ cm, gluino masses are excluded, when combining LEP~\cite{Baer} and CDF data, from 3 to at least $120$ GeV. 
          
We now consider the analysis for $\lambda_T=38$ cm. We consider the ($l2g1p1$), ($l2g1p2$) and ($l2g2p3$) scenarios. 

The case ($l2g1p1$) is plotted in Fig.\ \ref{f:equal38}. Larger $\lambda_T$
implies fewer R-hadron collisions with nuclei. This means that they generally move faster inside the hadronic calorimeter, because they are slowed down less by hadronic collisions with the nuclei. The fast moving charged particles lose less energy due to ionization than slower particles. Thus there is less energy deposited in the ($l2g1p1$)
scenario than ($l1g1p1$). This significantly increases the number of
muon-like jets as compared to ($l1g1p1$) because the jets are more likely to deposit less than $6$ GeV transverse energy and they are identified muon-like if they satisfy the other requirements. The gluino mass is excluded up to slightly lower than $120$ GeV.  The signal rate also decreases substantially below $50$ GeV. We observed only one event at $35$ GeV. We thus conclude that we can not exclude a gluino mass below $35$ GeV
in our analysis. {\em We therefore suggest that there is an open window between $\sim 22-25$ GeV and $35$ GeV requiring further exploration.}      

In the ($l2g1p2$) scenario (see Fig.\ \ref{f:eqr338}) we observe smaller signal cross sections than ($l2g1p1$) for the same reason that ($l1g1p2$) has smaller signal cross sections than ($l1g1p1$). This is due to the greater number of events that contain muon-like jets. The gluino mass is excluded up to $\sim 115$ GeV. At small gluino mass we should not expect a significant difference between (l2g1p1) and ($l2g1p2$) as mentioned before. Therefore, there is again an open window between $\sim 22-25$ GeV and $35$ GeV in this case. 

The ($l2g2p3$) scenario is plotted in Fig.\ \ref{f:100r038}. There are
no muon-like events in this scenario and we expect a larger signal rate compared
to ($l2g1p1$).   We find events even at a gluino mass of $35$ GeV and thus there is no open window below $35$ GeV.  The Gluino mass is excluded up to $\sim 125$ GeV.

\subsection{Lowering the squark masses}

To examine the effect of squarks on the results, we decrease the squark masses. A colliding quark and anti-quark make a virtual gluon in the s-channel which turns into a gluino pair.  They can also exchange a squark in the t-channel and produce a pair of gluinos.  When all squark masses are large, the former process dominates. Upon lowering the squark masses, the negative interference of the latter process decreases the gluino pair production cross section. However, the gluino production cross section due to the process $gg\to\tilde{g}\tilde{g}$ remains unchanged. When squark masses are sufficiently decreased then squarks can be produced directly in proton-antiproton collisions. We therefore have events containing a gluino and a squark jet. The gluino-squark production cross
section, however, is smaller than the gluino pair production cross section by a factor of approximately 100 at large gluino mass (around $140$ GeV) and a factor of 1000 at smaller gluino mass (around $60$ GeV). At large gluino masses, the signal rate increases substantially due to the squark production which enhances the jet signal rate and more than compensates the decrease in the total production cross section of gluino pairs. This however is not the case at smaller gluino masses. In fact the signal rate decreases upon lowering the squark masses (we have considered squarks as light as 450 GeV). At small gluino mass, squarks are produced at a relatively smaller rate than in the case of large gluino mass and the enhancement due to the squark signals are much less. The gluino pair production cross section decreases as mentioned before and the overall effect is that a lower signal rate is observed.  Recalling that we found an allowed window for gluinos between $\sim 22-25$ GeV and $35$ GeV in the case of $\lambda_T=38$ cm and equal probabilities for producing the four R-hadrons, {\em the decrease in the signal rate has the effect of enlarging the heavy gluino window, pushing it to slightly higher values of the gluino mass.}                      

\section{CONCLUSIONS}
\label{sec:conclude}
In this paper we considered a heavy gluino to be the lightest supersymmetric particle [LSP].  We used CDF Run I data, for excess events in the jets plus missing momentum channel, to place limits on the mass of a heavy gluino LSP. 
We find that gluinos with mass in the range between $\sim 35$ GeV and $\sim 115$ GeV are excluded.  Combined with the previous results of Baer et al. [BCG] ~\cite{Baer}, which use LEP data to exclude the range 3 - 22$\sim$25 GeV, {\em our result demonstrates that an allowed window for a heavy gluino with mass between 25 and 35 GeV is quite robust.}

We compared our analysis to that of BCG.  They also found an open window, however, for seemingly unphysical values of the fragmentation probability parameter $P \geq 3/4 $.  In their analysis, the bare gluino fragments into a {\em charged} R-hadron with the fragmentation probability $P$ and into a {\em neutral} R-hadron with probability $1 - P$.    These authors also assume that when an R-hadron scatters on a nucleon the light brown muck (quarks and gluons bound to the gluino) are stripped off.  The gluino then re-fragments with the same probability $P$.  Thus for BCG, the probability $P$ plays two independent roles:
\begin{itemize}
\item fragmentation probability, and
\item  rescattering probability.
\end{itemize}
In our analysis, however,  Regge scattering cross sections allow R-hadrons to change their identity with velocity dependent probabilities in each hadronic collision.  We thus separate the independent phenomena of fragmentation and rescattering.   As a result, we find a heavy gluino window with physical fragmentation probabilities discussed in sec. \ref{sec:prod}.

Both our results and those of BCG are sensitive to the R-hadron - Nucleon scattering length $\lambda_T$.   The heavy gluino window only exists for $\lambda_T = 38$ cm.   For the smaller value of $\lambda_T = 19$ cm, the window is closed; consistent with the results of BCG.  Although we do not know a priori the value of the scattering length, we argued that larger values are preferred.  

Note, as $\lambda_T$ is increased, at some point the R-hadron will no longer have hadronic collisions in the calorimeter.  The neutral R-hadron will escape the detector, while $\tilde \rho^{\pm}$ will only undergo ionization energy loss; behaving just like a heavy charged lepton.   We do not expect the heavy gluino window between 25 and 35 GeV (discussed here) to be significantly affected.  On the other hand, one might be concerned that searches for charged stable massive particles would now provide significant contraints on a heavy gluino in this limit.   However BCG have performed an analysis of charged stable massive particles (in the context of the heavy gluino LSP) using CDF data ~\cite{hoffman}.  They find that no limit can be set for a gluino mass less than 50 GeV.

 In addition, an open window requires significant R-hadron charge fluctuations.
We showed that in the case of purely neutral R-hadrons, gluinos are excluded with mass between 3 and 120 GeV (see Fig. \ref{f:100r038}).

Charge fluctuations are relevant for the analysis, since the CDF data cuts events with isolated leptons, in particular muons with $E_T > 10$ GeV.   A track is identified as a muon if it is charged in the central tracker and in at least one of the muon chambers. It's energy must also be more than 10 GeV, as measured in the central tracker, but less than 6 GeV in the hadronic calorimeter.  Thus R-hadrons can be identified as muons if they initially fragment charged and then end up in at least one of the muon chambers charged.   However in order to deposit less than 6 GeV in the hadronic calorimeter, it is important to have a longer scattering length (fewer hadronic collisions) and significant charge fluctuations (less ionization energy loss).   

It is clear from our analysis that gluinos with mass in the range 25 - 35 GeV (and $\lambda_T$ = 38 cm) are cut from the CDF sample by the muon cut and even more so by the 60 GeV missing momentum cuts.  This greatly diminished CDF's sensitivity to the heavy gluino LSP in this mass range.  It may be possible to re-analyze the Run I data with revised cuts which enhance the signal to background ratio.  Such an analysis is however beyond the scope of this paper.

Consider alternate methods for finding the heavy gluino.
We note that of order $ 10^6$ R-hadrons, with mass in the allowed window, were produced at CDF in Run I.  Assume only a few percent are stopped in the hadronic calorimeter.   These will be absorbed into the iron nuclei and thus a small sample of the calorimeter can be tested in a mass spectrometer to search for heavy isotopes of iron.   It is also possible for the gluino to be the next-to-lightest superparticle and decay into a gravitino and a gluon with lifetimes on the order of 100 years~\cite{Raby2}.   This possibility would not affect any of the analysis carried out in this paper.\footnote{Also, such an unstable R-hadron with a 100 year lifetime is still a good UHECRon candidate~\cite{Raby2}.}  On the other hand, this possibility would result in a spectacular heavy gluino signal.  The stopped R-hadrons would decay into a gravitino plus hadrons with visible hadronic energy $m_{R_0} - m_{Gravitino}$ which could be of order 10 GeV.  Clearly one can now look for events in the detector, when the accelerator is OFF.

In conclusion, it is important to search for a heavy gluino LSP with mass in the allowed range 25 - 35 GeV.   After all, this may be where the elusive SUSY is hidden.   In addition, as discussed by Albuquerque et al. these particles are the prime particle physics candidates for the so-called UHECRons~\cite{albuquerque}, responsible for the observed ultra high energy cosmic ray showers.

\section{ACKNOWLEDGMENTS}
 
This work was supported by the DOE/ER/01545-771.  We would also like to thank the Fermilab Theory Group for their hospitality and the partial support of a Frontier Fellowship for some of this work.  We thank H. Baer, E. Braaten, R. Hughes and K. Tobe for helpful discussions;  also S. Mrenna and K. Tobe for giving us a revised version of Spythia relevant for the gluino LSP model. We also benefited from discussions with R. Dermisek who investigated the case of a light gluino LSP.


\begin{thebibliography}{99}

\bibitem{sugra} A.H. Chamseddine, R. Arnowitt and P. Nath,  Phys. Rev. Lett. {\bf 29}, 970 (1982);  R. Barbieri, S. Ferrara and C.A. Savoy,  Phys. Lett. B {\bf 119}, 343 (1983);   L.J. Hall, J. Lykken and S. Weinberg,  Phys. Rev. D {\bf 22}, 2359 (1983);  P. Nath, R. Arnowitt and A.H. Chamseddine, 
 Nucl. Phys. B {\bf 322}, 121 (1983).


\bibitem{cmssm}  G.L. Kane, C. Kolda, L. Roszkowski and J.D. Wells, 
 Phys. Rev. D {\bf 49}, 6173 (1994).


\bibitem{gmsb}  S. Dimopoulos and S. Raby, Nucl. Phys. B {\bf 192}, 353 (1981);  M. Dine, W. Fischler, and M. Srednicki,  Nucl. Phys. B {\bf 189}, 575 (1981); M. Dine and W. Fischler, Phys. Lett. B {\bf 110}, 227 (1982);
M. Dine and M. Srednicki,  Nucl. Phys. B {\bf 202}, 238 (1982);
L. Alvarez-Gaum\'{e}, M. Claudson, and M. Wise,  Nucl. Phys. B {\bf 207}, 96 
(1982); C. Nappi and B. Ovrut,  Phys. Lett. B {\bf 113}, 175 (1982).


\bibitem{drw}  S. Dimopoulos, S. Raby and F. Wilczek,  Phys. Rev. D
{\bf 24}, 1681 (1981); S. Dimopoulos and H. Georgi,  Nucl.
Phys. B {\bf 193}, 150 (1981); L. Ibanez and G.G. Ross,  Phys.
Lett. B {\bf 105}, 439 (1981); M. B. Einhorn and D. R. T. Jones, 
Nucl. Phys. B {\bf 196}, 475 (1982); W. J. Marciano and G. Senjanovic,
 Phys. Rev. D {\bf 25}, 3092 (1982).  For  recent analyses, see P.
Langacker and N. Polonsky,  Phys. Rev. D {\bf 47}, 4028 (1993);
ibid., {\bf 49}, 1454 (1994); M. Carena, S. Pokorski, C.E.M. Wagner,
 Nucl. Phys. B {\bf 406}, 59 (1993).


\bibitem{Raby1} S. Raby, Phys. Rev. D {\bf 56}, 2852 (1997); S. Raby, Phys. Lett. B {\bf 422}, 158 (1998).

\bibitem{Raby2} S. Raby and  K. Tobe, Nucl. Phys. {\bf B539}, 3 (1999).

\bibitem{string} V. S. Kaplunovsky and J. Louis, Phys.
Lett. B {\bf 306}, 269 (1993); A. Brignole, L.E. Ibanez, C. Munoz,
Nucl. Phys. {\bf B422}, 125, (1994), {\bf B436}, 747(E), (1995); A. Brignole, L.E. Ibanez, C. Munoz, Report No. CERN-TH/97-143, hep-ph/9707209.  The fact
that this model can have a gluino LSP was discussed in the paper by  
C.H. Chen, M. Drees, J.F. Gunion,  Phys. Rev. D {\bf 55} 330 (1997), Erratum-ibid. D {\bf 60} 039901 (1999). 

\bibitem{nandi}  R.N. Mohapatra and S. Nandi, Phys. Rev. Lett. {\bf 79}, 181 (1997); Z. Chacko, B. Dutta, R.N. Mohapatra and S. Nandi, Phys. Rev. D {\bf 56}, 
5466 (1997).


\bibitem{Dimopoulos} S. Dimopoulos and F. Wilczek, Proceedings Erice Summer
School, Ed. A. Zichichi (1981).

\bibitem{babubarr}  K.S. Babu and S.M. Barr, Phys. Rev. D {\bf 48}, 5354 (1993);
V. Lucas and S. Raby,  Phys. Rev. D {\bf 55}, 6986 (1997).

\bibitem{cosmological} R.N. Mohapatra and V.L. Teplitz, Phys. Rev. Lett. {\bf 81}, 3079 (1998);  A.E. Faraggi, K.A. Olive and M. Pospelov, (hep-ph/9906345).

\bibitem{Baer} H. Baer, K. Cheung and J. F. Gunion,
Phys. Rev. D {\bf 59}, 075002 (1999).


\bibitem{Farrar} There are many papers on light gluinos. For example, see
G. Farrar and P. Fayet,  Phys. Lett. B {\bf 76}, 575 (1978); R. Barbieri, 
L. Girardello and A. Masiero,  Phys. Lett.  B {\bf 127}, 429  (1983); L. Clavelli,  Phys. Rev. D {\bf 45}, 3276, (1992); ibid.,  {\bf 46}, 2112 (1992); L. Clavelli et al., Phys. Rev. D {\bf 47}, 1973 (1993); M. Jezabek and J.H. 
K\"{u}hn,  Phys. Lett. B {\bf 301}, 121 (1993);  J. Ellis, D.V. Nanopoulos 
and D.A. Ross,  Phys. Lett. B {\bf 305}, 375  (1993); G.R. Farrar and A. Masiero, RU-94-38 (hep-ph/9410401) (1994); L. Clavelli and P.W. Coulter,  Phys. Rev. D {\bf 51}, 1117 (1995);  G. R. Farrar, Phys. Rev. D {\bf 51}, 3904 (1995); ibid., Phys. Rev. Lett. {\bf 76}, 4111 (1996); ibid., {\bf 76}, 4115 (1996);
L. Roszkowski and M. Shifman,  Phys. Rev. D {\bf 53}, 404 (1996).


\bibitem{chung}  D.J.H. Chung, G.R. Farrar and E.W. Kolb,  Phys. Rev. D {\bf 57}, 4606 (1998).

\bibitem{albuquerque}  I.F.M. Albuquerque, G.R. Farrar and E.W. Kolb, Phys. Rev. D {\bf 59}, 015021 (1999).

\bibitem{ktev}  KTev Collaboration (A. Alavi-Harati et al.), Phys. Rev. Lett. {\bf 83}, 2128 (1999).


\bibitem{hewett} J.L. Hewett,  T.G. Rizzo and M.A. Doncheski, Phys. Rev. D {\bf 56}, 5703 (1997).




\bibitem{Csikor} F. Csikor and Z. Fodor, Phys. Rev. Lett. {\bf 78},
4335 (1997); Report No. ITP-BUDAPEST-538, hep-ph/9712269; Z. Nagy
and Z. Trocsanyi, hep-ph/9708343; Phys. Rev. D {\bf 57}, 5793 (1998);
KTeV Collaboration, J. Adams {\it et al.}, Phys. Rev. Lett. {\bf 79},
4083 (1997); DELPHI Collaboration, P. Abreu {\it et al.}, Phys. Lett. B
{\bf 414}, 410 (1997); ALEPH Collaboration, R. Barate {\it et al.}, Z. Phys.
C {\bf 96}, 1 (1997) 

\bibitem{Farrar1} G. R. Farrar, hep-ph/9707467.

\bibitem{Starkman} G.D. Starkman, A. Gould, R. Esmailzadeh and
S. Dimopoulos, Phys. Rev. D {\bf 41}, 3594 (1990).

\bibitem{Dover} C.B. Dover, T.K. Gaisser and G. Steigman,
Phys. Rev. Lett. {\bf 42}, 1117 (1979).

\bibitem{Wolfram} S. Wolfram, Phys. Lett. B {\bf 82}, 65 (1979).

\bibitem{Mohapatra} R.N. Mohapatra and S. Nussinov, Phys.
Rev. D {\bf 57}, 1940 (1997).

\bibitem{otherlab}  R.N. Mohapatra, F. Olness, R. Stroynowski and V.L. Teplitz, (hep-ph/9906421);  C.S. Li, P. Nadolsky, C.P. Yuan and H.Y. Zhou, Phys. Rev. D {\bf 58}, 095004 (1999).


\bibitem{Gunion} J.F. Gunion, preprint UCD­98­2, Proceedings 
of the International Workshop on Quantum Effects in the MSSM,
UAB, Barcelona, September 9--13, 1997, ed. J. Sola (World Scientific Publishing).



\bibitem{opal}  OPAL Collaboration, Phys. Lett. B {\bf 377}, 273 (1996).

\bibitem{Exper} CDF Collaboration, J. Hauser, in {\it Proceedings of the 10th
Topical Workshop on Proton-Antiproton Collider Physics} (AIP, New York, 1995).

\bibitem{chanowitz}  M.S. Chanowitz, Phys. Rev. D {\bf 12}, 918 (1975). 

\bibitem{Drees} M. Drees, X. Tata, Phys. Lett. B, {\bf 252}, 695 (1990).


\bibitem{Gunion1} J. F. Gunion and D. E. Soper, Phys. Rev. D
{\bf 15}, 2617 (1977).


\bibitem{Chew} G.F. Chew, S.C. Frautschi, Phys. Rev. Lett. {\bf 7}, 394, (1961);
Phys. Rev. Lett. {\bf 8}, 41, (1962). 

\bibitem{Barger} V. D. Barger, D. B. Cline, {\it Phenomenological
Theories of High Energy Scattering, An Experimental Evaluation},
W. A. Benjamin, Inc., (1969).

\bibitem{Mueller} A. H. Mueller, Phys. Rev. D {\bf 2}, 2963, (1970).

\bibitem{Field} R. D. Field and  G. C. Fox, Nucl. Phys. {\bf B80}, 367 (1974).

\bibitem{spythia} S. Mrenna, Comput. Phys. Commun. {\bf 101}, 232 (1997).


\bibitem{Data} Particle Data Group, R. M. Barnette {\it et al.}, Review of
Particle Properties, Phys. Rev. D, {\bf 54}, 1 (1996)

\bibitem{hoffman}  K. Hoffman,  Report No. FERMILAB-CONF-97/430-3, http://www-cdf.fnal.gov/physics/exotic/conference/conference.html.


\end{thebibliography}
\end{document}